 \documentclass[final,1p,times,twocolumn,authoryear]{elsarticle}
 \biboptions{numbers,sort&compress}
 \bibpunct{[}{]}{,}{n}{}{,}
\usepackage{amssymb}
\usepackage{lipsum}
\usepackage{bm}
\usepackage{yfonts}
\usepackage{color}

\usepackage{graphicx,epsfig}% Include figure files
\usepackage{amsmath,amssymb,mathtools}%,amsfonts,multirow,rotate,color}amssymb
\usepackage{braket}
\def\be{\begin{equation}}
\def\ee{\end{equation}}

\def\bc{\begin{center}}
\def\ec{\end{center}}
\def\bea{\begin{eqnarray}}
\def\eea{\end{eqnarray}}

\newcommand{\Avg}[1]{\left\langle{#1}\right\rangle}

\journal{Physics Letters B}

\begin{document}

\begin{frontmatter}

\title{$\Lambda$-Gravity from Entropy }

\author[first]{Ginestra Bianconi}
\affiliation[first]{organization={School of Mathematical Sciences, Queen Mary University of London},
            addressline={Mile End Road}, 
            city={London},
            postcode={E1 4NS}, 
            country={United Kingdom}}

\begin{abstract}
The Gravity from Entropy (GfE) theory treats the metric of spacetime as quantum operators and proposes the GfE action for gravity. The GfE Lagrangian is given by the Geometric Quantum Relative Entropy (GQRE) between the true metric and the metric induced by the matter fields and curvature. The theory leads to modified gravity equations and predicts the emergence of a dynamical cosmological constant $\Lambda_{\mathcal{G}}$.  In the low energy, small curvature limit the GfE gravity equations reduce to Einstein equations with vanishing cosmological constant, i.e. in the weak field linear approximation $\Lambda_{\mathcal{G}}$ vanishes. In this work, we propose a generalisation of the GfE theory, called $\Lambda$-Gravity from Entropy ($\Lambda$GfE). Following a thermodynamic approach, we introduce a Lagrange multiplier $\Lambda$ to define the local volume of the theory.  {By enforcing local  {Weyl stationarity} of the $\Lambda$GfE action modulo the volume constraint of the theory,  we derive the equation for  $\Lambda$. The  $\Lambda$GfE equations of motion  include two dark energy contributions: the  emergent and dynamical contribution given by $\Lambda_{\mathcal{G}}$ plus a constant contribution given by $\Lambda$ where $\Lambda$  can be identified with the cosmological constant in the low energy small curvature limit.} 
 {While this theory cannot predict the exact value of $\Lambda$, here it is shown that  the vacuum contributions to the dressed stress–energy tensor cancel  from the trace-free   equations of motion of the  $\Lambda$GfE theory.  Within the GfE framework, the $\Lambda$GfE also offers a novel interpretation of the dark energy terms, with potentially testable consequences for cosmological observations.}
\end{abstract}




\end{frontmatter}

%\tableofcontents



\section{Introduction}

Symmetries are a pillar of physical theories and a guiding principle for fundamental physics. Local symmetries play a central role, as exemplified by the local Lorentz invariance of General Relativity (GR).  Considerable interest has been devoted to conformal symmetry and local conformal symmetry in theoretical high-energy physics and gravity~\cite{t2015local,mannheim2012making,bars2014local,capozziello2010review,shaposhnikov2023anomaly}. Notably, while local conformal invariance is a symmetry of Maxwell’s equations in vacuum,  the Einstein–Hilbert action does not exhibit local conformal symmetry. Although conformal (Weyl) gravity~\cite{mannheim2012making,maldacena2011einstein} has been proposed as a viable modified gravity theory, it is widely believed that local conformal symmetry of spacetime cannot be an exact symmetry for generic quantum theories due to the  Weyl anomaly~\cite{birrell1984quantum}.

Despite the important conceptual challenges involved in arguing for a local conformal symmetry of gravity, ’t Hooft suggested in an inspiring paper~\cite{t2015local} that formulating a modified gravity theory with this symmetry might shed new light on the cosmological constant problem. The cosmological constant problem~\cite{weinberg1989cosmological,carroll2001cosmological} is often framed in terms  of  the identification of the cosmological constant with the vacuum energy of quantum field theories—an expectation that has been strongly challenged by the large discrepancy between its observed value and the theoretical predictions based on this assumption. Although several mechanisms to address this problem have been proposed in the literature~\cite{weinberg1989cosmological,carroll2001cosmological,copeland2006dynamics,capozziello2013cosmographic}, the cosmological constant problem remains one of the most outstanding problems in theoretical physics. It is often argued  that some yet-unclear mechanism must be responsible for cancelling the enormous vacuum energy contributions while still yielding the small but non-zero observed value of the cosmological constant.
In this context, unimodular theory~\cite{smolin2009quantization,eichhorn2013unimodular,de2021can,finkelstein2001unimodular,unruh1989unimodular}, dating back to an original proposal by Einstein~\cite{einstein1915feldgleichungen},   {together with its generally covariant and Weyl-transverse formulations~\cite{henneaux1989cosmological,oda2017classical,alonso2025spacetime}, and trace-free general relativity equations \cite{ellis2011trace,ellis2014trace} } have  recently been revisited and proposed as a viable alternative to address the cosmological constant problem. This research direction has been argued \cite{smolin2009quantization,ellis2011trace} to handle  the cancellation of the contributions coming from the trace of the stress-energy tensor, also named  the first cosmological constant problem using the terminology of Ref. \cite{smolin2009quantization}.

 {In this paper, we propose the $\Lambda$-Gravity from Entropy ($\Lambda$GfE) theory. This theory enforces  {Weyl stationarity}  modulo the volume constraint of the theory and opens new perspectives on the cosmological constant problem within the GfE theory.  } Our starting point is the recently proposed Gravity from Entropy (GfE) ~\cite{bianconi2025gravity,bianconi2025quantum,bianconi2025thermodynamics} which provides a new theoretical physics approach for addressing the relation between gravity, statistical mechanics, and quantum information theory that plays a fundamental role in physics, as  evidenced by the thermodynamics of black holes and de Sitter spaces~\cite{bekenstein1973black,hawking1975particle,bunch1978quantum,kubizvnak2017black,clifton2013gravitational}.  While  entropic gravity approaches base their arguments on  the entropy of holographic screens and horizons, \cite{jacobson1995thermodynamics,jacobson2016entanglement,verlinde2011origin,carroll2016entropy,dorau2025quantum,yokokura2025black}
the GfE theory~\cite{bianconi2025gravity,bianconi2025quantum,bianconi2025thermodynamics} postulates that gravity is fundamentally a statistical mechanics theory of the metric degrees of freedom. Specifically, the GfE theory~\cite{bianconi2025gravity,bianconi2025quantum,bianconi2025thermodynamics,thattarampilly2025inflation,pereira2026vacuum,thattarampilly2026spherically} treats the metrics associated with spacetime as quantum operators and proposes the GfE action for gravity. The  Lagrangian of the GfE theory is given by the Geometric Quantum Relative Entropy (GQRE) between the true metric of the manifold and the metric induced by the matter fields and the curvature. The GfE theory reduces to GR with zero cosmological constant in the limit of low energy and small curvature, but beyond this limit proposes a theory of modified gravity depending on an emergent field called G-field. Thus, the modified GfE equations can lead to testable predictions. The full GfE theory includes an emergent dark energy contribution $\Lambda_{\mathcal{G}}$ that is always positive and is dependent on the G-field which can be interpreted as the internal energy density of the GfE theory~\cite{bianconi2025thermodynamics}.

In cosmology, it has been shown that the GfE theory leads to a new mechanism for inflation without the need to introduce additional scalar fields \cite{thattarampilly2025inflation} which can have potential consequences for the Early Universe while the weak field corrections to general relativity have been investigated in Ref.  \cite{thattarampilly2026spherically}.
The emergent dark energy contribution $\Lambda_{\mathcal{G}}$ of the GfE theory is, however, vanishing in the low energy and small curvature limit, thus the de Sitter space can only be recovered if one assumes that the stress-energy tensor has a non-zero vacuum energy contribution.

In order to address this problem, here we propose the $\Lambda$GfE theory.
Motivated by the thermodynamic interpretation of the GfE theory proposed recently \cite{bianconi2025thermodynamics},   we introduce a new constraint in the GfE Lagrangian, in addition to the constraint determined by the G-field. This new constraint fixes the value of the local volume $\omega$, associated with the local GfE entropy, to $\sqrt{-|g|}$. As we will discuss in the following, the Lagrange multiplier of this constraint, indicated as  $\Lambda$, provides a dark energy contribution to the $\Lambda$GfE theory in addition to the emergent dark energy contribution $\Lambda_{\mathcal{G}}$.   
%We note that  the actual value of the local volume $\omega$ is not fixed a priori, as in unimodular theory, rather, it is determined solely by the constraint enforcing its value to be $\sqrt{-|g|}$. %This implies that we are not allowed to consider the variation of this generalized action with respect to $\omega$ as $\omega$ is not to be considered an independent scalar density of weight plus one. 
 The introduced constraint appears to be, to some extent, redundant, as it vanishes on shell.
However, we note that this seemingly innocuous term leads to an additional contribution to the cosmological constant of modified gravity equations proportional to $\Lambda$. 
 {A key conceptual step of the $\Lambda$GfE theory is  to determine $\Lambda$ by imposing  {Weyl stationarity}  of the entire $\Lambda$GfE action modulo the volume constraint of the theory. The $\Lambda$GfE action comprises the original GfE action, which is not locally conformal invariant taken in isolation, together with the new addition to the action enforcing the mentioned constraint which allows  {the  Weyl stationarity principle} to hold.}

As it happens for unimodular theory \cite{smolin2009quantization} and trace-free Einstein equations \cite{ellis2011trace,ellis2014trace},  the $\Lambda$GfE theory  leads  to  modified gravity equations in which we observe a cancellation of the trace of the stress-energy tensor contributions. However, the resulting overall contribution to dark energy includes both a constant term, given by $\Lambda$, and a dynamical term $\Lambda_{\mathcal{G}}$ dependent on the G-field, which is subleading in the weak field limit.

Therefore, the $\Lambda$GfE theory in the low energy, small curvature limit displays the de Sitter limit in which $\Lambda$ is to be identified with the cosmological constant, but beyond this limit it includes dynamical corrections to the cosmological constant.   {Thus, this work provides additional elements of relevance to  the existing literature on  GfE cosmology \cite{thattarampilly2025inflation,bianconi2025thermodynamics}, and might open new perspectives for the present and future experimental observations of the Hubble tension \cite{di2025cosmoverse,karim2025desi,carloni2025does} and other astrophysical observations \cite{berti2015testing,barack2019black}.}

  \section{ Gravity from Entropy (GfE) theory}
  \subsection{The GfE action}
  We consider a  $d=4$ dimensional spacetime with signature $(-1,1,1,1)$ given by a torsion-free pseudo-Riemannian manifold $\mathcal{K}$ associated with the metric $g_{\mu\nu}$ and its Levi-Civita connection.  Throughout this work, we take units $c=\hbar=1$ and we denote the Planck length by $\ell_P$.
  The true higher-order metric associated with $\mathcal{K}$ is given by the direct sum of a scalar identity, the true metric between vectors and the true metric between bivectors, i.e.
 \bea
\tilde{g}=1\oplus g_{\mu\nu}dx^{\mu}\otimes dx^{\nu}\oplus [g_{(2)}]_{\mu\nu\rho\sigma}dx^{\mu}\wedge dx^{\nu}\otimes dx^{\rho}\wedge dx^{\sigma},
\eea
with 
\bea
[g_{(2)}]_{\mu\nu\rho\sigma}=\frac{1}{2}(g_{\mu\rho}g_{\nu\sigma}-g_{\mu\sigma}g_{\nu\rho}).
\eea
In addition to the true higher-order metric we associate to spacetime the higher-order metric $\tilde{\bf G}$ induced by the matter fields and the curvature given by
\bea
\tilde{\bf G}=G_{(0)}\oplus [G_{(1)}]_{\mu\nu}dx^{\mu}\otimes dx^{\nu}\oplus [G_{(2)}]_{\mu\nu\rho\sigma}dx^{\mu}\wedge dx^{\nu}\otimes dx^{\rho}\wedge dx^{\sigma}.
\eea
This metric depends on the connection and on the curvature of spacetime whose higher-order expression $\tilde{\bm{\mathcal{R}}}$ is the direct sum of the Ricci scalar, the Ricci tensor and the Riemann tensor, i.e.
\bea
\tilde{\bm{\mathcal{R}}}=a_0 R\oplus  a_1R_{\mu\nu}dx^{\mu}\otimes dx^{\nu}\oplus R_{\mu\nu\rho\sigma}dx^{\mu}\wedge dx^{\nu}\otimes dx^{\rho}\wedge dx^{\sigma},
\eea
where $a_0\geq 0,a_1\geq 0$ are dimensionless parameters, here and in the following taken to be equal to one, i.e. $a_0=a_1=1$. 
The GfE theory postulates that the true metric and the metric induced by the matter fields and the curvature can be treated as quantum operators that operate on the Hilbert space defined by the matter fields as outlined in Ref. \cite{bianconi2025gravity}. By interpreting the metric induced by the matter fields and curvature as a {\it renormalizable density matrix}, the GfE theory defines the induced metric $\tilde{\bf G}$  as 
\bea
\tilde{\bf G}={\bf I}+\alpha \tilde{\bf M}-\beta\tilde{\bm{\mathcal{R}}},
\eea
where $\tilde{\bf M}$ indicates the contribution of the matter fields and is given by 
\bea
\tilde{\bf M}&=&M_{(0)}\oplus [M_{(1)}]_{\mu\nu}dx^{\mu}\otimes dx^{\nu}\oplus [M_{(2)}]_{\mu\nu\rho\sigma}dx^{\mu}\wedge dx^{\nu}\otimes dx^{\rho}\wedge dx^{\sigma},
\eea 
while the parameters $\alpha$ and $\beta$ are given by $\alpha=\alpha^{\prime}\ell_P^4, \beta=\beta^{\prime}\ell_P^2$ with $\alpha^{\prime},\beta^{\prime}$ dimensionless constants.
 {Here and in the following we will consider a scalar matter field $\phi$ associated with the choice \cite{bianconi2025gravity}
\bea
M_{(0)}=m^2\bar{\phi}\phi,\quad [M_{(1)}]_{\mu\nu}=\nabla_{\mu}\bar{\phi}\nabla_{\nu}\phi,\quad [M_{(2)}]_{\mu\nu\rho\sigma}=0.
\eea
For an explicit generalized expression of $\tilde{\bf M}$ associated with Dirac-K\"ahler bosonic fields and Abelian gauge fields see Ref. \cite{bianconi2025gravity}; while for the expression of $\tilde{\bf M}$ valid for perfect fluids see Ref.  \cite{bianconi2025thermodynamics}.} 
The GfE theory \cite{bianconi2025gravity} proposes for gravity the GfE  action $\mathcal{S}_{\textrm{GfE}}$  given by 
\bea
\mathcal{S}_{\textrm{GfE}}=\frac{1}{\ell_P^4}\int\sqrt{-|g|}\mathcal{L}_{\textrm{GfE}}d^4{\bf r},
\eea
where the GfE Lagrangian $\mathcal{L}_{\textrm{GfE}}$ \cite{bianconi2025gravity} is given by the Geometrical Quantum Relative Entropy (GQRE) between the true higher-order metric of spacetime and the higher-order metric induced by the matter fields and curvature, i.e.
\bea
\mathcal{L}_{\textrm{GfE}}=-\mbox{Tr}_F\ln({\bf \tilde{G}}\tilde{g}^{-1}),
\label{LGfE}
\eea
where $\mbox{Tr}_F$ indicates the flattened trace defined in Ref. \cite{bianconi2025gravity}.
Interestingly, the GfE action admits a Boltzmann interpretation and  can be shown to  quantify the local degrees of freedom of the matter fields and the geometry. In fact,  it can be expressed in a Boltzmann-like entropy form~\cite{bianconi2025quantum,bianconi2025thermodynamics}
\bea
\mathcal{L}_{\textrm{GfE}}=\ln W,
\eea
where $W$ is given by
\bea
W=G_{(0)}^{-1}(\mbox{det}(g_{(1)}{\bf G}_{(1)}^{-1}))(\mbox{det}(g_{(2)}{\bf G}_{(2)}^{-1})).
\eea
It was recently realized that the GfE theory is associated with the GfE thermodynamics \cite{bianconi2025thermodynamics}  in which the local entropy $s$  is given by 
\bea
s=\omega\mathcal{L}_{\textrm{GfE}} 
\eea
where the {\em local volume} $\omega$ is given  {on shell} by the determinant of the metric, i.e. 
\bea
\omega=\sqrt{-|g|}.
\eea
This scalar density of weight plus one therefore admits a thermodynamic interpretation and plays an important role in the first principle of GfE thermodynamics derived in Ref. {\cite{bianconi2025thermodynamics}}. As we will discuss in the following, this quantity will play a fundamental role in the $\Lambda$GfE theory as well. 
\subsection{The G-field and the emergent dark energy contribution $\Lambda_{\mathcal{G}}$}
The  GfE Lagrangian given by Eq.(\ref{LGfE}) is non-linear in the curvature, however this non-linear dependence involves  exclusively  the quantity 
\bea
\tilde{\bm \Theta}=\tilde{\bf G}\tilde{g}^{-1}.
\label{Gtilde}
\eea
 Thus, the GfE Lagrangian can be expressed by enforcing these constraints with Lagrange multipliers $\tilde{\bm{\mathcal{G}}}$ leading to the following expression for the Lagrangian \cite{bianconi2025gravity}
\bea
{\mathcal{L}}_{\textrm{GfE}}=-\mbox{Tr}_F\ln{\tilde{\bm \Theta}}-\mbox{Tr}_F\left[\tilde{\bm{\mathcal{ G}}}(\tilde{\bf G}\tilde{g}^{-1}-\tilde{\bm\Theta })\right].
\label{L2GfE}
\eea
In the GfE theory  the G-field $\tilde{\bm{\mathcal{G}}}$ is treated as a physical and possibly measurable field, which can be related to $\tilde{\bm\Theta}$ by
\bea
\bm{\mathcal{G}}=\tilde{\bm\Theta}^{-1}=[\tilde{\bf G}\tilde{g}^{-1}]^{-1},
\label{Gfield}
\eea
where in the last expression we used Eq.(\ref{Gtilde}).
Thus, the GfE theory assumes that the G-field can be treated like other notable Lagrange multipliers  in statistical mechanics which are physical and measurable, such as the temperature and the chemical potential. 
Substituting Eq.(\ref{Gfield}) into Eq.(\ref{L2GfE}) we obtain that the Lagrangian  decouples into a gravitational Lagrangian $\mathcal{L}_G$ and a matter Lagrangian  $\mathcal{L}_M$, i.e.
\bea
{\mathcal{L}}_{\textrm{GfE}}=\beta\mathcal{L}_G+\alpha\mathcal{L}_M
\eea
The gravitational  Lagrangian and the matter Lagrangian involve contraction with a {\em dressed metric} $\tilde{g}_{\mathcal{G}}$ given by 
\bea
\tilde{g}_{\mathcal{G}}={\tilde{\bm{\mathcal{G}}}}^{-1}\tilde{g}.
\eea
Moreover, the GfE Lagrangian ${\mathcal{L}}_{\textrm{GfE}}$ includes also an {\em emergent dynamical cosmological constant} $\Lambda_{\mathcal{G}}$ that depends exclusively on the G-field 
\bea
\Lambda_{\mathcal{G}}=\frac{1}{6\beta}\mbox{Tr}_F\left[\tilde{\bm{\mathcal{G}}}-\tilde{\bf I}-\ln \tilde{\bm{\mathcal{G}}}\right],
\label{LambdaG}
\eea
where $\tilde{\bf I}$ indicates the higher-order identity (see Ref.  \cite{bianconi2025gravity} for details).
Specifically, the gravitational  Lagrangian $\mathcal{L}_G$ and the matter Lagrangian $\mathcal{L}_M$  are given by 
\bea
\mathcal{L}_G= \left({\mathcal{R}}_{\mathcal{G}}-6\Lambda_{\mathcal{G}}\right),\quad \mathcal{L}_M=- \mbox{Tr}_F \tilde{{\bf M}}\tilde{g}_{\mathcal{G}}^{-1},\label{LGM}
\eea
where the dressed Ricci scalar ${\mathcal{R}}_{\mathcal{G}}$ is given by
\bea
{\mathcal{R}}_{\mathcal{G}}=\mbox{Tr}_F \tilde{\bm{\mathcal{R}}}\tilde{g}_{\mathcal{G}}^{-1}.
\eea
From this discussion it follows that  giving a physical interpretation to the G-field provides a reformulation of the GfE action  that is linear in the curvature and the G-field and allows us to cast the GfE action into the sum of a gravitational Lagrangian and a matter Lagrangian. Note, however, that the presence of the G-field in both Lagrangians  implies that matter and geometry cannot be treated independently.
The GfE action is diffeomorphism invariant as a whole, however, we note that the GfE theory, like GR, which constitutes its low energy small curvature limit,  is not locally conformal invariant. 
\subsection{The GfE gravity equations}
Imposing that the variation of the  GfE action $\mathcal{S}_{\textrm{GfE}}$ with respect to the metric $g^{\mu\nu}$ vanishes, %i.e. imposing 
%\bea
%\frac{\delta S_{\textrm{GfE}}}{\delta g^{\mu\nu}}=0
%\eea 
leads to the GfE  gravity equations  for the metric ${g}_{\mu\nu}$  given by \cite{bianconi2025gravity}
\bea
{R}^{\mathcal{G}}_{(\mu\nu)}-\frac{1}{2}{g}_{\mu\nu}\Big(\mathcal{R}_{\mathcal{G}}-6\Lambda_{\mathcal{G}}\Big)+{{\mathcal{D}}}_{(\mu\nu)}=\frac{\alpha}{2\beta}\mathcal{T}_{\mu\nu},
\label{GfEeq}
\eea
where  $(\mu\nu)$ indicates the symmetrisation of the indices, ${R}^{\mathcal{G}}_{\mu\nu}$ are the elements or the {\em dressed Ricci tensor} given by
\bea
{R}^{\mathcal{G}}_{\mu\nu}&=&{\mathcal{G}_{(0)}}R_{\mu\nu}+{[{\mathcal{G}_{(1)}}]}_{\mu}^{\ \rho}R_{\rho\nu}-{[\mathcal{G}_{(2)}]}_{\rho_1\rho_2\mu\eta}R_{\nu}^{\ \eta\rho_1\rho_2}\nonumber \\
&&+2{[{\mathcal{G}_{(2)}}]}_{\mu}^{\ \eta\rho_1\rho_2}R_{\rho_1\rho_2\nu\eta},
\eea
while ${{\mathcal{D}}}_{\mu\nu}$ are the elements depending on second derivatives of the G-field $\tilde{\bm{\mathcal{G}}}$ given by 
\bea
{{\mathcal{D}}}_{\mu\nu}&=&(\nabla^{\rho}\nabla_{\rho}g_{\mu\nu}-\nabla_{\mu}\nabla_{\nu}){\mathcal{G}_{(0)}}-\nabla^{\rho}\nabla_{\nu}{[\mathcal{G}_{(1)}]}_{(\rho\mu)}
\nonumber \\
&&+\frac{1}{2}
\nabla^{\rho}\nabla_{\rho}{[\mathcal{G}_{(1)}]}_{\mu\nu}+\frac{1}{2}\nabla^{\rho}\nabla^{\eta}{[\mathcal{G}_{(1)}]}_{\rho\eta}g_{\mu\nu}\nonumber \\
&&+\nabla^{\eta}\nabla^{\rho}{[\mathcal{G}_{(2)}]}_{\mu\rho\nu\eta}+\nabla^{\rho}\nabla^{\eta}{[\mathcal{G}_{(2)}]}_{\eta\mu\rho\nu}\nonumber \\
&&+\frac{1}{2}[\nabla^{\rho},\nabla^{\eta}]{[\mathcal{G}_{(2)}]}_{\rho\eta\mu\nu}.
\eea
Defining the matter action $\mathcal{S}_M$ as 
\bea
\mathcal{S}_M=\int\sqrt{-|g|}\mathcal{L}_M d^4{\bf r},
\label{SM}
\eea
the {\em dressed stress-energy tensor} $\mathcal{T}_{\mu\nu}$ appearing in the GfE equations of motion (\ref{GfEeq}) is given by
\bea
\mathcal{T}_{\mu\nu}=\frac{-2}{\sqrt{-|g|}}\frac{\delta \mathcal{S}_M}{\delta g^{\mu\nu}}.
\eea
The investigation of the exact solutions of the GfE equations of motion has the potential to shed new light on cosmology and has been argued in Ref.  \cite{thattarampilly2025inflation} to lead to inflationary behaviour without the need to introduce a scalar field.

\subsection{The GR limit of the GfE theory}
In the  limit of low energy and small curvature the GfE equations of motion reduce to the Einstein equations with zero cosmological constant. This can be seen by expanding the GfE Lagrangian $\mathcal{L}_{\textrm{GfE}}$ for  low energy and small curvature obtaining, up to a multiplicative constant,   the GR Lagrangian $\mathcal{L}_{GR}$, i.e. 
\bea
\mathcal{L}_{\textrm{GfE}}\simeq \alpha\mathcal{L}_{GR}=\alpha\left[\frac{1}{2\kappa}R+\mathcal{L}_M\right]
\label{GRlimit}
\eea
where  $\kappa$ is the Einstein gravitational constant and the matter field Lagrangian has the usual definition $ \mathcal{L}_M=-\mbox{Tr}_F\tilde{\bf M}\tilde{g}^{-1}$.
Note that in this derivation we have  constrained the GfE parameters $\alpha$ and $\beta$ to satisfy
${\alpha}/({3\beta})=2\kappa$.
The equations of motion of the GfE theory in the low energy small curvature limit are therefore 
the GR equations with zero  cosmological constant
\bea
R_{\mu\nu}-\frac{1}{2}g_{\mu\nu}R=\kappa T_{\mu\nu}.
\eea
It follows that, unless we identify the cosmological constant with the energy density of the vacuum,  in the low energy, small curvature limit, the GfE equations of motion do not include the de Sitter space.
%There is no cosmological constant, so the de Sitter space is not a solution of the modified equations as long as one does not consider the addition of a vacuum energy inserted ``by hand" as a matter field.
%inserting the constant this is too small respect to the measured once

\section{The  Weyl stationarity  principle and the $\Lambda$GfE theory}
\subsection{General framework}
The geometric degrees of freedom of the GfE theory are not only captured by the interplay between matter,  curvature and geometry, through the GQRE depending on the G-field Eq.(\ref{Gfield}), but depend also on the local volume of the  theory $\omega$ identified  {on shell} with the determinant of the metric ${\sqrt{-|g|}}$.
In analogy with  our introduction of the G-field through Lagrange multipliers, we therefore introduce the volume $\omega$ of the theory by introducing the Lagrange multiplier $\Lambda$ which enforces the constraint 
 {\bea
C[\omega, g_{\mu\nu}]=\omega-{\sqrt{-|g|}}=0,
\label{v_constraint}
\eea}
 {where $\omega$ and $\sqrt{-|g|}$ are scalar densities of weight plus one.}
 {In the $\Lambda$GfE theory $\omega$ is therefore introduced as an auxiliary constrained quantity, not as an Euler-Lagrange field.}
Specifically,  in this theory $\omega$ is not known a priori and thus it does not constrain the true metric, as in unimodular theory. Rather,  Eq.(\ref{v_constraint}) is used to define $\omega$  starting from the determinant of the metric ${\sqrt{-|g|}}$. Therefore, all ten degrees of freedom of the metric $g_{\mu\nu}$ are dynamical.
Thus, introducing the Lagrange multiplier $\Lambda$, enforcing the constraint Eq.(\ref{v_constraint}),  we obtain the $\Lambda$GfE Lagrangian ${\mathcal{L}}_{\Lambda\textrm{GfE}}$ given by
\bea
{\mathcal{L}}_{\Lambda\textrm{GfE}}=\mathcal{L}_{\textrm{GfE}}+\mathcal{L}_{\Lambda},\eea 
where $\mathcal{L}_{\textrm{GfE}}$  follows Eq. (\ref{L2GfE}) and $\mathcal{L}_{\Lambda}$ reads 
\bea
\mathcal{L}_{\Lambda}&=&\frac{6\beta\Lambda}{\omega}\left(\omega-\sqrt{-|g|}\right).
\eea
It follows that the $\Lambda$GfE action is given by
 \bea
 \mathcal{S}_{\Lambda\textrm{GfE}}=\mathcal{S}_{\textrm{GfE}}+\mathcal{S}_{\Lambda},
 \label{LGfE_action}
 \eea
 with 
 \bea
\mathcal{S}_{\textrm{GfE}}&=&\frac{1}{\ell_P^4}\int \sqrt{-|g|}\mathcal{L}_{\textrm{GfE}} d^4{\bf r},\nonumber \\
\mathcal{S}_{\Lambda}&=&\frac{1}{\ell_P^4}\int \sqrt{-|g|}\mathcal{L}_{\Lambda} d^4{\bf r}.
 \eea
 Note that the action $S_{\textrm{GfE}}$ depends on all components of the curvature decomposition. As we mentioned before, taken in isolation, this action is diffeomorphism invariant, while it is not locally conformal invariant.
The constraint Eq.(\ref{v_constraint}) enforcing the definition of the local volume $\omega$ can be simply obtained from the action $S_{\Lambda\textrm{GfE}}$ defined in Eq.(\ref{LGfE_action}) by imposing that the variation of the Lagrangian with respect to $\Lambda$ vanishes.

 {If we were to treat $\omega$ as an Euler–Lagrange field, the value of $\Lambda$ would be determined by requiring the variation of the action with respect to $\omega$, at fixed $g_{\mu\nu}$, to vanish.
  {Thus in this way we would obtain either $\sqrt{-|g|}=0$ which is unphysical, 
or $\Lambda=0$   which clearly leaves the GfE equations of motion unchanged as expected. This latter option would indeed only correspond to a change of variables.}}

 {Instead, in the $\Lambda${GfE} theory, we do not treat $\omega$ as an Euler-Lagrange field. Rather,    we  determine  $\Lambda$ by imposing {\em  {Weyl stationarity} modulo the volume constraint} of the $\Lambda$GfE theory. To this end, we consider infinitesimal Weyl transformations of the Euler-Lagrange fields $g_{\mu\nu},\phi$ and $\tilde{\bm{\mathcal{G}}}$ while $\omega$ is not treated as an independently varied field. 
 We require the corresponding first variation of the $\Lambda$GfE action to vanish upon restriction to the constraint surface $C[\omega, g_{\mu\nu}]=\omega-\sqrt{-|g|}=0$.}
 {This principle will  allow us to introduce  the $\Lambda$GfE theory demonstrating that  the introduction of the seemingly innocuous term $S_{\Lambda}$  modifies the equations of motion. This Weyl stationarity principle modulo the volume constraint, should be distinguished from a conventional Weyl gauge symmetry acting tangentially on the full constrained configuration space.}

 {To illustrate the Weyl stationarity principle modulo the volume constraint let us consider the variation of the Euler-Lagrange fields under  local conformal transformations of the metric, given by}
\bea
 g_{\mu\nu}&\to &e^{2\sigma}{g}_{\mu\nu},\quad \phi\to e^{-\Delta_{\phi}\sigma}\phi,\quad \tilde{\bm{\mathcal{G}}}\to e^{-\Delta_{\mathcal{G}}\sigma}\tilde{\bm{\mathcal{G}}}
 \label{local_conf}
\eea
where $\Delta_{\phi}$ and $\Delta_{\mathcal{G}}$ are the conformal weights for the matter field $\phi$ and the G-field $\tilde{\bm{\mathcal{G}}}$ respectively.
  {In order to impose  {Weyl stationarity} modulo the volume constraint, }
 we impose that the variation $ \delta_{\sigma}\mathcal{S}_{\Lambda\textrm{GfE}}$ of the $\Lambda$GfE action with respect to $\sigma$  vanishes modulo the volume constraint imposed by the Lagrange multiplier  $\Lambda$,  {i.e. on the manifold $C[\omega,g_{\mu\nu}]=0$, where ${\delta\mathcal{S}_{\Lambda\textrm{GfE}}}/{\delta \Lambda}=0$ we have}
 { \bea
 \delta_{\sigma}\mathcal{S}_{\Lambda\textrm{GfE}}=\frac{\delta\mathcal{S}_{\Lambda\textrm{GfE}}}{\delta \Lambda}\delta_{\sigma}\Lambda=0.\label{modulo}
\eea}
Effectively, this procedure determines  $\Lambda$ by imposing  {Weyl stationarity} modulo the volume constraint, leading to a generalisation of the GfE theory which we call $\Lambda$GfE theory. 
Following this approach, we obtain (see \ref{ApA} for details)
 {\bea
\left.2\frac{\delta \mathcal{S}_{\textrm{GfE}}}{\delta g_{\mu\nu}}g_{\mu\nu}-\frac{\delta \mathcal{S}_{\textrm{GfE}}}{\delta \bm{\mathcal{G}}}\Delta_{\mathcal{G}}\bm{\mathcal{G}}-\frac{\delta \mathcal{S}_{\textrm{GfE}}}{\delta\phi}\Delta_{\phi} \phi+
2\frac{\delta \mathcal{S}_{\Lambda}}{\delta g_{\mu\nu}}g_{\mu\nu}\right|_{C[\omega,g_{\mu\nu}]=0}=0.
\label{new}
\eea}

 {Interestingly, we observe that this {Weyl stationarity principle,} treating $\omega$ as an auxiliary constrained variable  does not imply that the constraint $C[\omega,g_{\mu\nu}]=\omega-\sqrt{-|g|}=0$ is itself Weyl invariant.  Indeed, under local conformal transformations the constraint $C[\omega,g_{\mu\nu}]$ transforms as \bea
C[\omega,g_{\mu\nu}]=\omega-\sqrt{-|g|}=0,\to C[\omega,e^{2\sigma}g_{\mu\nu}]=\omega-e^{4\sigma}\sqrt{-|g|}=0.\eea
Thus it is this constraint itself that enforces that {\em after the transformation, and  when the constraint is satisfied}, $\omega=e^{4\sigma}\sqrt{-|g|}$.} {It follows that the formulation of the $\Lambda$GfE theory remains generally covariant, since the volume constraint is imposed covariantly by imposing the constraint $C[\omega,g_{\mu\nu}]=0$, rather than by fixing $\sqrt{-g}$ to a  constant. This is analogous in spirit to covariant unimodular gravity, while differing from Weyl-transverse formulations in which local Weyl transformations are part of the gauge symmetry \cite{henneaux1989cosmological,oda2017classical,alonso2025spacetime}. In the present construction, Weyl stationarity should therefore not be identified a priori with an independent gauge redundancy.}

Anticipating the results of the next section, we will also derive the equations of motion of the $\Lambda${GfE} theory by further imposing that  the variation of the action $S_{\Lambda\textrm{GfE}}$ with respect to ${g}_{\mu\nu}$, with respect to the matter field $\phi$ and with respect to the G-field $\tilde{\bm{\mathcal{G}}}$ vanishes. These equations of motion differ from the equations of the GfE theory as they contain two dark energy terms: $\Lambda$ and $\Lambda_{\mathcal{G}}$.
Interestingly, by observing that $\mathcal{S}_{\textrm{GfE}}$ is diffeomorphism invariant, we will show that on shell the Lagrange multiplier $\Lambda$ is constant. However, this treatment will not be able to predict the exact value of $\Lambda$.

%\end{document}
  {We observe that in GR there are other scenarios in which the introduction of a constraint in the Lagrangian  can modify the equations of motion. Specifically, this phenomenology occurs  in the treatment of perfect matter fluids (with $p=0$) in cosmology.   Indeed, the Lagrangian for a perfect matter fluid is $\mathcal{L}_{PF}=\rho(u_\mu u^{\mu}+1)$ which can be interpreted as a constraint imposing  the norm of the four-velocity $u_{\mu}$ of the fluid to be minus one, i.e. $u_\mu u^{\mu}=-1$. Notably, this Lagrangian modifies the equations of motion of GR, when the  constraint is satisfied, leading to the stress-energy contribution $T_{\mu\nu}=\rho u_{\mu}u_{\nu}$. In this derivation, the unconstrained variation with respect to $u_{\mu}$ is not considered physically relevant as (i) this Lagrangian should be complemented with extra terms including the particle number current $J_{\mu}$ and other fields and constraints meant to determine $u_{\mu}$ and (ii) the variation of $u_{\mu}$ is assumed to lie on the manifold  orthogonal to the constraint, thus $u_{\mu}\delta u^{\mu}=0$. 
 
 Similarly,  in the $\Lambda$GfE theory we do not consider physically relevant the variation of $\mathcal{S}_{\Lambda}$ with respect to $\omega$, assuming that (i)  the full quantum gravity $\Lambda$GfE theory  will be required to  determine the exact value of  $\omega$ and  that (ii) $\omega$ can be interpreted  as the determinant of a metric $\hat{g}$, i.e. $\omega \equiv \sqrt{-|\hat{g}|}$,  with the variation of $\hat{g}$ constrained to lie on the manifold orthogonal to the constraint in Eq.$(\ref{v_constraint})$, i.e. orthogonal to conformal transformations.  A working hypothesis is that the putative determination of  $\omega$ in the full quantum $\Lambda$GfE might be important to determine the exact value of the cosmological constant. Furthermore, drawing a direct parallel with the perfect matter fluid example,  we emphasize that, as previously mentioned,  while $\mathcal{S}_{\Lambda}$ vanishes when the constraint is satisfied, the equations of motion of $\Lambda$GfE differ from the equations of motion of GfE. Notably, as we will see in the following, they  include an additional and constant dark energy contribution $\Lambda$.}
 
In order to illustrate the consequence of the $\Lambda$GfE theory let us first consider its low energy, small curvature limit which will allow us to reveal the relation between $\Lambda$ and the cosmological constant.
\subsection{The $\Lambda$GfE theory in the low energy small curvature limit} 
 In the low energy small curvature limit, the $\Lambda$GfE theory reduces to the trace-free gravity equations where $\Lambda$ plays the role of the cosmological constant term.
 Specifically, we obtain, for low energy and small curvature,
 \bea
\mathcal{L}_{\Lambda\textrm{GfE}}&\simeq &\alpha\mathcal{L}_{GR}+\mathcal{L}_{\Lambda}\eea
where we have used Eq.(\ref{GRlimit}) to approximate the GfE Lagrangian $\mathcal{L}_{\textrm{GfE}}$.
By imposing that the variation with respect to $\Lambda$ vanishes, we recover the constraint in Eq.(\ref{v_constraint}). Imposing that $\mathcal{S}_{\Lambda\textrm{GfE}}$ obeys  {the Weyl stationarity principle  modulo the volume constraint}, i.e. imposing Eq.(\ref{modulo}) we obtain Eq. (\ref{new}) and thus, as long as the matter field and the G-field obey the equations of motion {(see \ref{ApA} for the details),}
 \bea
 \Lambda=\frac{R}{4}+ \kappa\frac{T}{4},
 \label{Lambda_GR}
 \eea
 where we have indicated  $T$  the trace of the stress-energy tensor, i.e. $T=T_{\mu}^{\ \mu}$.
 The variation of the action $S_{\Lambda\textrm{GfE}}$ with respect to the metric $g_{\mu\nu}$ leads, in the low energy, small curvature limit to the $\Lambda$GfE equations of motion
\bea
  R_{\mu\nu}-\frac{1}{2}g_{\mu\nu}\left(R-2\Lambda\right)=\kappa T_{\mu\nu} 
 \eea
 Therefore, using Eq.(\ref{Lambda_GR}) the $\Lambda$GfE equations of motion reduce to the trace-free Einstein equations \cite{ellis2011trace,ellis2014trace} in the low energy small curvature limit, i.e.
 \bea
 R_{\mu\nu}-\frac{1}{4}g_{\mu\nu}R=\kappa \left(T_{\mu\nu}-\frac{1}{4}Tg_{\mu\nu}\right) 
 \eea
  These equations are known to solve the vacuum energy problem as it is apparent that any contributions of the type $\rho_{\textrm{vac}}g_{\mu\nu}$ to the stress-energy tensor do not contribute to these equations.
  At the same time, however, the equations recover all the vacuum solutions of the Einstein equations unchanged,  so e.g.  Schwarzschild and Kerr black hole solutions.
 These results demonstrate that  the $\Lambda$GfE theory shares with unimodular theory the advantage that the vacuum energy contributions to the stress-energy tensor constitutively cancel due to the presence of the $\Lambda$ term \cite{smolin2009quantization,ellis2011trace,ellis2014trace}. Note, however, that the $\Lambda$GfE theory is distinct from unimodular theory because $\omega$ is not fixed a priori, but effectively determined by imposing the  {Weyl stationarity principle.}
 
The $\Lambda$GfE gravity equations also imply that $\Lambda$ is a constant. In fact, noticing that $\mathcal{L}_{M}$ is diffeomorphism invariant and the Ricci curvature $R$ satisfies the Bianchi identity we have, on shell  {(see \ref{ApB} for the details),}
\bea
&&\nabla^{\mu}\left[R_{\mu\nu}-\frac{1}{2}Rg_{\mu\nu}\right]=0,\quad\nabla^{\mu}T_{\mu\nu}=0,\label{cons_GR}
\eea   
which implies that on shell, we must have
\bea
\nabla^{\mu}\Lambda=0.
\eea
Thus $\Lambda$ is constant on shell and can be identified with the  cosmological constant.
Using Eq.(\ref{Lambda_GR}) this equation imposes the constraint
\bea
\nabla^{\mu}\left[\frac{R}{4}+\kappa \frac{T}{4}\right]=0.
\eea

Summing up, the $\Lambda$GfE theory in the low energy small curvature limit leads to trace-free Einstein equations  and therefore implies that the   {metric-proportional} vacuum fluctuations of the matter field do not contribute to equations of motion as in unimodular theory  \cite{smolin2009quantization,ellis2011trace,ellis2014trace}.
However, it is well known \cite{smolin2009quantization} that these trace-free Einstein equations are not able to predict the actual value of the cosmological constant, which is of  order $10^{-120}/\ell_P^2$. 
  {The present $\Lambda$GfE theory, while not solving this challenging  problem, points out that a possible key to interpreting this experimental value might come from a full quantum $\Lambda$GfE theory.}
%\end{document}
\subsection{The equations of motion of the $\Lambda$GfE theory}
Moving beyond the low energy small curvature limit, the equations of motion of the $\Lambda$GfE can be obtained by imposing   {\em   {Weyl stationarity}} of the $\Lambda$GfE theory modulo the volume constraint, thus obtaining that $\Lambda$ is given by Eq.(\ref{new}). As long as the matter fields and the G-field follow the equations of motion, we obtain  {(see \ref{ApA} for the details),}  
\bea
\Lambda=\frac{1}{12}\mathcal{R}_{\mathcal{G}}-\frac{1}{12}{{\mathcal{D}}}+\frac{1}{4}\kappa\mathcal{T}-\Lambda_{\mathcal{G}},
\label{Lambda_GfE}
\eea
where ${{\mathcal{D}}}$ indicates the trace of ${{\mathcal{D}}}_{(\mu\nu)}$.
Furthermore, by imposing that the variation of the action $S_{\Lambda\textrm{GfE}}$  with respect to $g_{\mu\nu}$ vanishes, one obtains the $\Lambda$GfE equations of motion 
\bea
{R}^{\mathcal{G}}_{(\mu\nu)}-\frac{1}{2}{g}_{\mu\nu}\Big(\mathcal{R}_{\mathcal{G}}-6\Lambda_{\mathcal{G}}-6\Lambda\Big)+{{\mathcal{D}}}_{(\mu\nu)}=3\kappa\mathcal{T}_{\mu\nu},
\label{LGfEeq}
\eea
where we have imposed the constraints in Eq. (\ref{v_constraint}).
With respect to the original GfE equation of motion, these equations of motion contain two dark energy terms: the first one depends on the G-field and is proportional to the emergent dark energy contribution $\Lambda_{\mathcal{G}}$ which was already present in the original GfE equations,  the second one is instead proportional to the new term $\Lambda$ of the $\Lambda$GfE theory.

Using the expression of $\Lambda$ that enforces  {Weyl stationarity} given by Eq.(\ref{Lambda_GfE}) the $\Lambda$GfE gravity equations can be equivalently expressed in trace-free form as 
 {\bea
\left[{R}^{\mathcal{G}}_{(\mu\nu)}-\frac{1}{4}{g}_{\mu\nu}\mathcal{R}_{\mathcal{G}}\right]+\left[{{\mathcal{D}}}_{(\mu\nu)}-\frac{1}{4}{{\mathcal{D}}}g_{\mu\nu}\right]=3\kappa\left[\mathcal{T}_{\mu\nu}-\frac{1}{4}g_{\mu\nu}\mathcal{T}\right].
\label{tfLGfEeq}
\eea}
 As already discussed for the trace-free equations derived in the low energy small curvature limit, also in this more general scenario the $\Lambda$GfE equations of motion  are not affected by   {metric-proportional vacuum fluctuations} of the matter fields.  {Indeed, vacuum fluctuations of the dressed stress-energy tensor $\mathcal{T}_{\mu\nu}$ can be shown (see \ref{ApC}) to  provide an effective contribution  proportional to the metric, i.e. $-\rho_{\textrm{vac}}g_{\mu\nu}$.}  This result  is derived in \ref{ApC} in the semiclassical regime neglecting the back-reactions on the metric $g_{\mu\nu}$ and on the G-field $\bm{\mathcal{G}}$ which can only be  fully addressed by a full quantum gravity formulation of $\Lambda$GfE.   
 Clearly,  this contribution to the dressed stress-energy tensor due to vacuum fluctuations cancels out and does not contribute to the equations of motion of the $\Lambda$GfE theory as the theory possesses  trace-free equations of motion. 
 {We note, however, that this cancellation of metric-proportional vacuum contributions should not be confused with radiative stability of the full quantum theory, as discussed for unimodular gravity \cite{padilla2015note}, or with vacuum-energy sequestering mechanisms based on additional global constraints \cite{kaloper2014vacuum}.}

Since $\mathcal{S}_{\textrm{GfE}}$ is  invariant under diffeomorphisms,  as $\mathcal{L}_{\textrm{GfE}}$ is a scalar, we have that   {on shell of the $\Lambda$GfE theory (see \ref{ApB} for details)}
\bea
\nabla^{\mu}\left[{R}^{\mathcal{G}}_{(\mu\nu)}-\frac{1}{2}{g}_{\mu\nu}\Big(\mathcal{R}_{\mathcal{G}}-6\Lambda_{\mathcal{G}}\Big)+{{\mathcal{D}}}_{(\mu\nu)}-3\kappa\mathcal{T}_{\mu\nu}\right]=0.
\label{cons_E}
\eea
This indicates that the stress-energy tensor is not conserved in isolation. Rather,  the presence of a non-trivial G-field $\tilde{\bm{\mathcal{G}}}$ that deviates from the identity couples the geometric degrees of freedom with the matter fields in this joint conservation law. Interestingly, this conservation law is dependent on the emergent dark energy term $\Lambda_{\mathcal{G}}$ that is not explicitly present in the trace-free equations of motion Eq.(\ref{tfLGfEeq}).
Given the above conservation law, Eq.(\ref{cons_E}),  using the equations  of motion of the $\Lambda$GfE theory  (\ref{LGfEeq}), we obtain that,  {on shell of the $\Lambda$GfE theory,} $\Lambda$ is  constant, i.e.
\bea
\nabla^{\mu}\Lambda=0.\eea
This also implies that while the emergent dark energy term $\Lambda_{\mathcal{G}}$ does not appear anymore in the trace-free expression of the $\Lambda$GfE equations of motion (\ref{tfLGfEeq}), it instead plays a critical role in determining  this  conservation principle  as we have  \bea
\nabla^{\mu}\Lambda_{\mathcal{G}}=\nabla^{\mu}\left[\frac{1}{12}\mathcal{R}_{\mathcal{G}}-\frac{1}{12}{{\mathcal{D}}}+\frac{1}{4}\kappa\mathcal{T}\right].
\eea
In summary, the $\Lambda$GfE equations of motion Eq.(\ref{LGfEeq}), by enforcing the  {Weyl stationarity principle}, lead to the inclusion of two dark energy contributions. The first is dependent on the G-field and is proportional to the emergent and dynamical $\Lambda_{\mathcal{G}}$ contribution. In the low energy small curvature limit this contribution vanishes but the term diverges in the high energy large curvature limit.  {The second term, which allows the $\Lambda$GfE theory to obey the  {Weyl stationarity principle} modulo the volume constraint, is proportional to $\Lambda$ which is constant on shell and plays the  role of the cosmological constant. }

\section{Conclusions}

 {In this work, we have proposed the $\Lambda$GfE theory that builds and extends the original results of the GfE theory by  imposing    {Weyl stationarity} modulo the volume constraint of the theory.
The basic idea is to introduce in the GfE Lagrangian  an apparently innocuous term  that is used to define the local volume of the theory. This additional term  is proportional to the Lagrange multiplier $\Lambda$. However, the introduction of this term allows us to impose  {the   {Weyl stationarity}} of the $\Lambda$GfE action,} modulo the volume constraint, a stationarity principle  that is not present in the original GfE action. This  {  {Weyl stationarity}} of the $\Lambda$GfE theory determines  the value of the Lagrange multiplier  $\Lambda$ which plays the role of a  dark energy contribution in addition to the dynamical dark energy contribution $\Lambda_{\mathcal{G}}$ already present in the GfE theory.

Interestingly, the $\Lambda$GfE equations of motion  reduce to  the trace-free Einstein equations where $\Lambda$ plays the role of the cosmological constant, in the low energy, small curvature limit.
These equations, like the trace-free Einstein equations,  are not affected by vacuum energy density contributions to the  {dressed} stress-energy tensor.  Beyond the low energy, small curvature limit the equations of motion include two dark energy contributions: a constant term given by $\Lambda$ and an emergent dynamical term $\Lambda_{\mathcal{G}}$ that depends on the G-field and is subleading in the weak field limit. 

 {These results offer a new perspective on both the cosmological constant problem and the broader quest for a quantum theory of gravity.  {Future work could establish whether,} on the theoretical side, the principle of  {\em   {Weyl stationarity} modulo the volume constraint}, proposed here within the $\Lambda$GfE theory,  may provide a useful guiding principle for exploring possible approaches to its second quantization.  {On the phenomenological side, future research could investigate whether the interplay between the two distinct contributions to the dark energy sector of the theory  may provide a new framework for interpreting the cosmological implications of the $\Lambda$GfE theory. These might  include the characterization of the  inflationary behaviour of the $\Lambda$GfE theory which might enrich and complement previous inflationary results obtained within  the GfE theory \cite{thattarampilly2025inflation}, and the possible exploration of the potential of the $\Lambda$GfE  to  provide interpretative clues for the cosmological observations related to the Hubble tension.}

\section*{Acknowledgements}
This publication was made possible through the support of grant \#63793 from the John Templeton Foundation.
This research benefited from discussions at the workshops "Quantum Universe" (Avellino, Italy, 2025) and  “Concepts of Quantum and Spacetime (KEK, Japan, March 2026).  {The author acknowledges relevant comments and remarks from an anonymous referee.} 
\bibliographystyle{elsarticle-num}

\bibliography{references}

\appendix

 {\section{ {Weyl stationarity} of the $\Lambda$GfE action modulo the volume constraint}
\label{ApA}
We consider the $\Lambda$GfE action 
\bea
{\mathcal{S}}_{\Lambda\textrm{GfE}}=\frac{1}{\ell_P^4}\int d^4x \sqrt{-|g|}\mathcal{L}_{\Lambda\textrm{GfE}}
\eea
where the $\Lambda$GfE action depends on the true metric $g_{\mu\nu}$, on the generic matter field $\phi$ and on the G-field $\bm{\mathcal{G}}$  and is given by 
\bea
{\mathcal{L}}_{\Lambda\textrm{GfE}}=\beta\mathcal{L}_G(g_{\mu\nu},\bm{\mathcal{G}})+\alpha\mathcal{L}_M(g_{\mu\nu},\bm{\mathcal{G}},\phi)+{\mathcal{L}}_{\Lambda}(g_{\mu\nu})
\eea
where
\bea
\mathcal{L}_G= \left({\mathcal{R}}_{\mathcal{G}}-6\Lambda_{\mathcal{G}}\right),\quad \mathcal{L}_M=- \mbox{Tr}_F \tilde{{\bf M}}\tilde{g}_{\mathcal{G}}^{-1},\quad
\mathcal{L}_{\Lambda}=\frac{6\beta\Lambda}{\omega}\left(\omega-\sqrt{-|g|}\right).
\label{LGM2}
\eea
For future convenience, we define $\mathcal{S}_G$ and $\mathcal{S}_M$ as 
\bea
\mathcal{S}_G=\int\sqrt{-|g|}\mathcal{L}_G d^4{\bf r},\quad \mathcal{S}_M=\int\sqrt{-|g|}\mathcal{L}_M d^4{\bf r}.
\label{SG_SM}
\eea

Under local conformal transformations given by Eq.(\ref{local_conf}), the $\Lambda$GfE action $\mathcal{S}_{\Lambda\textrm{GfE}}$ transforms as 
\bea
\delta_{\sigma} \mathcal{S}_{\Lambda\textrm{GfE}}&=&\int dx^4 \sqrt{-|g|}\left[2\frac{1}{\sqrt{-|g|}}\frac{\delta \mathcal{S}_{\Lambda\textrm{GfE}}}{\delta g_{\mu\nu}}g_{\mu\nu}\delta\sigma+ \frac{1}{\sqrt{-|g|}}\frac{\delta \mathcal{S}_{\Lambda\textrm{GfE}}}{\delta \phi}\delta_{\sigma} \phi\right.\nonumber \\&&
\left.+\frac{1}{\sqrt{-|g|}}\frac{\delta \mathcal{S}_{\Lambda\textrm{GfE}}}{\delta \bm{\mathcal{G}}}\delta_{\sigma}\bm{\mathcal{G}}+\frac{1}{\sqrt{-|g|}}\frac{\delta \mathcal{S}_{\Lambda\textrm{GfE}}}{\delta \Lambda}\delta_{\sigma}\Lambda\right].
\eea
Note that, as discussed in the main text,  $\omega$ (i) is not interpreted as an independent dynamical variable and (ii) when considering $\omega$ as the determinant of a metric $\hat{g}$, (i.e. $\omega\equiv\sqrt{-|\hat{g}|}$) only variations of the auxiliary metric $\hat{g}$ orthogonal to the constraint defined in Eq.(\ref{v_constraint}), (i.e. with $\delta_{\sigma}\hat{g}=0$) are considered.

Imposing  {Weyl stationarity} modulo the volume constraint 
\bea
\frac{\delta \mathcal{S}_{\Lambda\textrm{GfE}}}{\delta \Lambda}=0, 
\eea
implying $C[\omega,g_{\mu\nu}]=0$,
consists  of postulating 
 { \bea
\int dx^4 \sqrt{-|g|}\left[2\frac{1}{\sqrt{-|g|}}\frac{\delta \mathcal{S}_{\textrm{GfE}}}{\delta g_{\mu\nu}}g_{\mu\nu} \sigma+ \frac{1}{\sqrt{-|g|}}\frac{\delta \mathcal{S}_{\Lambda\textrm{GfE}}}{\delta \phi}\delta_{\sigma} \phi \right.\nonumber \\
+\left.\left.\frac{1}{\sqrt{-|g|}}\frac{\delta \mathcal{S}_{\Lambda\textrm{GfE}}}{\delta \bm{\mathcal{G}}} \delta_{\sigma}{\bm{\mathcal{G}}}+2\frac{1}{\sqrt{-|g|}}\frac{\delta \mathcal{S}_{\Lambda}}{\delta g_{\mu\nu}}g_{\mu\nu} \sigma \right|_{C[\omega,g_{\mu\nu}]=0}\right]=0,
\eea 
}
from which Eq.(\ref{new}) follows. When the matter field $\phi$ and the G-field obey the equations of motion, we have
\bea
\frac{\delta \mathcal{S}_{\Lambda\textrm{GfE}}}{\delta \phi}=0,\quad\frac{\delta \mathcal{S}_{\Lambda\textrm{GfE}}}{\delta \bm{\mathcal{G}}}=0.
\eea
Taking into account that the variation of  $\mathcal{S}_{\Lambda}$ is given by
 {\bea
\hspace{-8mm}\frac{1}{\sqrt{-|g|}}\left.\frac{\delta \mathcal{S}_{\Lambda}}{\delta g_{\mu\nu}}\right|_{C[\omega,g_{\mu\nu}]=0}&=&\left.\frac{3\Lambda \beta}{\ell_P^4}\frac{\sqrt{-|g|}}{\omega}\right|_{C[\omega,g_{\mu\nu}]=0}g^{\mu\nu}-\left.\frac{3\beta\Lambda}{\ell_P^4}\left(1-\frac{\sqrt{-|g|}}{\omega}\right)\right|_{C[\omega,g_{\mu\nu}]=0}g^{\mu\nu}\nonumber \\
&=&\frac{3\Lambda \beta}{\ell_P^4}g^{\mu\nu}
\eea
}
we recover, in the general scenario Eq.(\ref{Lambda_GfE})  and in the low energy, small curvature limit Eq.(\ref{Lambda_GR}).}

\section{Derivation of the conservation law due to the diffeomorphism invariance of the GfE action}
\label{ApB}
Let us now exclusively consider the GfE action 
\bea
{\mathcal{S}}_{\textrm{GfE}} = \frac{1}{\ell_P^4}\int d^4x \sqrt{-|g|}\mathcal{L}_{\textrm{GfE}}
\eea
which is associated with the Lagrangian ${\mathcal{L}}_{\textrm{GfE}}$ given by
\bea
{\mathcal{L}}_{\textrm{GfE}}=\beta\mathcal{L}_G(g_{\mu\nu},\bm{\mathcal{G}})+\alpha\mathcal{L}_M(\phi_i,g_{\mu\nu},\bm{\mathcal{G}}).
\eea
Since ${\mathcal{L}}_{\textrm{GfE}}$ is a scalar this action is diffeomorphism invariant.
By assuming that the considered diffeomorphism transformation $\xi^{\mu}$ vanishes sufficiently strongly at the boundary, we neglect the boundary terms and find that the variation of the GfE action ${\mathcal{S}}_{\textrm{GfE}}$ with respect to  $\xi^{\mu}$ is given by 
\bea
\delta_{\xi} \mathcal{S}_{\textrm{GfE}}&=&\int d^4x \sqrt{-|g|}\left[\frac{1}{\sqrt{-|g|}}\frac{\delta \mathcal{S}_{\textrm{GfE}}}{\delta g_{\mu\nu}}\delta_{\xi}g_{\mu\nu}+ \frac{1}{\sqrt{-|g|}}\frac{\delta \mathcal{S}_{\textrm{GfE}}}{\delta \phi}\delta_{\xi} \phi\right.\nonumber \\&&
+\left.\frac{1}{\sqrt{-|g|}}\frac{\delta \mathcal{S}_{\textrm{GfE}}}{\delta \bm{\mathcal{G}}}\delta_{\xi}\bm{\mathcal{G}}\right]
\eea
Since we have 
\bea
\frac{\delta \mathcal{S}_{\textrm{GfE}}}{\delta \phi}=\frac{\delta \mathcal{S}_{\Lambda\textrm{GfE}}}{\delta \phi},\quad 
\frac{\delta \mathcal{S}_{\textrm{GfE}}}{\delta \bm{\mathcal{G}}}=\frac{\delta \mathcal{S}_{\Lambda\textrm{GfE}}}{\delta \bm{\mathcal{G}}}.
\eea
If we are on shell of the $\Lambda$GfE theory we have 
\bea
\frac{\delta \mathcal{S}_{\textrm{GfE}}}{\delta \phi}=0,\quad
\frac{\delta \mathcal{S}_{\textrm{GfE}}}{\delta \bm{\mathcal{G}}}=0.
\eea
It follows that, under these conditions, the diffeomorphism invariance of the GfE theory implies that 
\bea
\delta_{\xi} \mathcal{S}_{\textrm{GfE}}&=&\int d^4x \sqrt{-|g|}\left[\frac{1}{\sqrt{-|g|}}\frac{\delta \mathcal{S}_{\textrm{GfE}}}{\delta g_{\mu\nu}}\delta_{\xi}g_{\mu\nu}\right]=0
\eea
The variation of the metric, $\delta_{\xi}g_{\mu\nu}$, is given by 
\bea
\delta_{\xi}g_{\mu\nu}=-2\nabla_{(\mu}\xi_{\nu)}
\eea
Thus, we obtain 
\bea
\delta_{\xi} \mathcal{S}_{\textrm{GfE}}&=&-2\int d^4x \sqrt{-|g|}\left[\frac{1}{\sqrt{-|g|}}\frac{\delta \mathcal{S}_{\textrm{GfE}}}{\delta g_{\mu\nu}}\nabla_{\mu}\xi_{\nu}\right]=0
\eea
which integrating by parts and neglecting boundary terms implies 
\bea
\delta_{\xi} \mathcal{S}_{\textrm{GfE}}&=&2\int d^4x \sqrt{-|g|}\left[\nabla_{\mu}\frac{1}{\sqrt{-|g|}}\frac{\delta \mathcal{S}_{\textrm{GfE}}}{\delta g_{\mu\nu}}\right]\xi_{\nu}=0
\eea
Thus this implies that on shell of the $\Lambda$GfE theory we have the conservation law,
\bea
\nabla_{\mu}\left[\frac{1}{\sqrt{-|g|}}\frac{\delta \mathcal{S}_{\textrm{GfE}}}{\delta g_{\mu\nu}}\right]=0,
\eea
which provides Eq.(\ref{cons_E}) in the general scenario and Eq.(\ref{cons_GR}) in the low energy, small curvature limit.
\section{Vacuum fluctuations in the semiclassical approach to $\Lambda$GfE}
\label{ApC}
 {In this Appendix we provide a semiclassical argument to show that in the framework of the $\Lambda$GfE    {the metric-proportional} vacuum energy density contributions  are captured by terms of the expectation of the {\em dressed} stress-energy tensor $\Avg{0_{\textrm{in}}\left|\mathcal{T}_{\mu\nu}\right|0_{\textrm{in}}}$ given by $-\rho_{\textrm{vac}}g_{\mu\nu}$. In order to discuss this result, let us consider a path integral approach starting from the matter partition function \bea
 Z=e^{-\textrm{i}W}=\int \mathcal{D}\phi e^{\textrm{i}\mathcal{S}_M} 
 \eea 
 where $\mathcal{S}_M$ is given by Eq.(\ref{SM}) with the matter Lagrangian given by Eq.(\ref{LGM}).
 It follows that the expectation of the dressed stress-energy tensor, in this semiclassical approach in which we consider both the true metric and the G-field as a fixed background, is given by
 \bea
 \Avg{\mathcal{T}_{\mu\nu}}=\frac{2}{\sqrt{-|g|}}\frac{\delta W}{\delta g^{\mu\nu}}.
 \eea
 {Restricting attention to the locally Lorentz-invariant metric-proportional vacuum-energy contribution, we write}  
  \bea
   \Avg{0_{\textrm{in}}\left|\mathcal{T}_{\mu\nu}\right|0_\textrm{in}}=-\rho_{\textrm{vac}}g_{\mu\nu}.
  \eea}
 

%% If you have bibdatabase file and want bibtex to generate thef
%\bibliographystyle{elsarticle-harv} 
%\bibliography{example}



%% For example:

%% \bibitem[Aladro et al.(2015)]{Aladro15} Aladro, R., Martín, S., Riquelme, D., et al. 2015, \aas, 579, A101

%%\end{thebibliography}

\end{document}